\documentstyle[prl,aps,twocolumn,epsf]{revtex}
\begin{document}
\draft
\tighten
\onecolumn
\twocolumn[\hsize\textwidth\columnwidth\hsize\csname
@twocolumnfalse\endcsname
\title{Electron spin coherence in semiconductors: Considerations for a
spin-based solid state quantum computer architecture}
\author{Rogerio de Sousa and S. Das Sarma} 
\address{Condensed Matter Theory Center, Department of Physics,
University of Maryland, College Park, MD 20742-4111} 
\date{\today} 
\maketitle
\begin{abstract}
We theoretically consider coherence times for spins in two quantum
computer architectures, where the qubit is the spin of an electron
bound to a P donor impurity in Si or within a GaAs quantum dot.  We
show that low temperature decoherence is dominated by spin-spin
interactions, through spectral diffusion and dipolar flip-flop
mechanisms.  These contributions lead to $1-100$ $\mu$s calculated
spin coherence times for a wide range of parameters, much higher than
former estimates based on $T_{2}^{*}$ measurements.
\end{abstract}
\vskip2pc]   
\narrowtext

Solid state quantum computation based on the intrinsic two level
dynamics of electron spin in semiconductors has attracted widespread
attention because the enormous resources of conventional electronics
can in principle be applied to develop a scalable quantum computer
(QC). In this context, solid state spins in an applied magnetic field
are attractive qubit candidates since they comprise perfect two-level
systems with potentially long coherence times. Nuclear spins of
phosphorus impurities in silicon are potential qubits because they are
well isolated from other degrees of freedom\cite{kane1}.  But for such
a quantum computer to work, precise electronic control of single
nuclear spins must be achieved, a rather daunting task.  Electron
spins are much easier to control. In that respect, electrons bound to
phosphorus impurities in silicon\cite{vrijen} and gallium arsenide
quantum dots\cite{loss1} are promising qubit candidates. However, for
quantum computation to be fault-tolerant these electron spins must be
coherent for at least $10^{4}$ elementary quantum
operations\cite{preskill},  which imposes a severe constraint since
very long spin coherence times  would be needed.  In this letter we
address the important question of principle involving  the fundamental
upper bound on electron spin coherence time in proposed  semiconductor
QC architectures.  Since such a device can only work at low
temperatures ($k_{B}T\ll E_{Z}$, where $E_{Z}$ is the Zeeman splitting
of the spins in an applied magnetic field B), we show that the
dominant  decoherence contribution comes from the unavoidable
spin-spin interactions with nuclei and other electrons.  This leads to
an unsurmountable upper bound on spin coherence, but our calculations
indicate that the fault-tolerance criterion can still be satisfied for
a wide range of parameters (e.g. quantum dot Fock-Darwin radius
$l_{0}$, concentration of $^{29}$Si isotopes $f$, {\it etc}) defining
the QC architecture. 

An essential property of electrons in Si:P and GaAs is that  their
electron spin resonance (ESR) line is inhomogeneously  broadened by
the hyperfine interaction with nuclear spins\cite{feher1,seck}. This
effect leads to a drastic difference between the precessing
magnetization of an ensemble of spins as compared to a single spin or
a group of them with the same Larmor frequency. The former decays in a
time scale $T_{2}^{*}$ which is dominated by the dephasing effect of
the inhomogeneous field distribution. The latter magnetization decays
in a time $T_{2}$ which is usually many orders of magnitude
longer\cite{revcoh}.  To extract the single spin coherence time
$T_{2}$ from an ESR experiment one has to perform a  $\pi/2-\pi$ spin
echo sequence\cite{mr_books}, that forces the spins to refocus
eliminating the dephasing effect.  This leads to a convenient
definition for spin coherence: Simply the time it takes for a spin
echo envelope to decay to 1/$e$ of its initial value.   Since spin
echoes usually decay in a quite different fashion from the simple
exponential predicted by the Bloch equations, we denote its decay time
by $T_{M}$ using a notation consistent with existing
literature\cite{mims}.  Spin echo experiments have been performed in
dilute Si:P\cite{gordon,chiba}, but never in GaAs. For the latter
material, claims of ``long coherence times'' have been based on
$T_{2}^{*}$ measurements only\cite{awsch_pt}, which do not reveal the
ultimate limit on spin coherence set by $T_{M}$, that can in principle
be many orders of magnitude longer than $T_{2}^{*}$.  Certainly
inhomogeneous broadening imposes severe tuning constraints on
one-qubit gates. But fortunately one can still have an universal QC
built only with two-qubit gates\cite{DiVincenzo00}. Our former work
shows that the spread of Zeeman frequencies will only weakly affect
two-qubit gates, since the exchange interaction is fairly insensitive
to inhomogeneous fields\cite{hu01}.  Therefore we emphasize that the
relevant spin coherence time for QC architectures is, in fact, $T_{M}$
(not $T_{2}^{*}$), and our theoretical finding of truly large values
of $T_{M}$ ($\gtrsim \mu$s) compared with the measured $T_{2}^{*}$
($\sim$ ns) values is quite significant.

Due to extreme sensitivity requirements, ESR in GaAs is usually
measured indirectly,  for example from the changes in
magneto-resistivity in a  2DEG\cite{magneto_resistance} or from
photoluminescence\cite{photo_lumi}. In these experiments one probes
ensemble spin properties of moving electrons or recombination pairs
respectively, meaning that one should be careful in extrapolating
those results to a single localized spin in a quantum computer
environment. Alternative methods to ESR include Faraday
rotation\cite{awsch_pt}, but again to study single spin coherence it
would be necessary to perform an echo sequence, that is yet to be done
using optical methods.  Measuring single spin coherence time in
semiconductors is a dauntingly difficult task, and therefore the need
for theoretical estimates of $T_{M}$ becomes acutely necessary.  Here
we present the first realistic calculation of $T_{M}$ for a GaAs
quantum dot (QD) based QC architecture (and also for Si:P QC
architecture). Our values exceed former coherence estimates based on
$T_{2}^{*}$ by three orders of magnitude, and establish quite
definitively that fault tolerant quantum computation should be
possible in semiconductor QC at low temperatures. 

First we discuss the spin-flip rate $T_{1}^{-1}$ with corresponding
energy transfer to the lattice.  Any spin-flip process contributes to
exponential decay of the spin echo signal. But bound electrons at low
temperatures quite generally have extremely long $T_{1}$ ($\gg
T_{M}$), since for the spin to flip, a virtual transition to an
excited orbital state must take place, quite different from the
conduction electron case where any momentum relaxation event may flip
the spin\cite{boundt1} through spin-orbit coupling. The direct phonon
emission rate for bound-electrons becomes $T_{1}^{-1}\propto
[n(E_{Z})+1]B^{5}$, where $n(E_{Z})=[\exp{(E_{Z}/k_{B}T)}-1]^{-1}$ is
the Bose occupation number for the emitted phonon wave-vector. Feher
and Gere measured $T_{1}^{-1}\approx 4\times 10^{-4}$ s$^{-1}$ for
Si:P at $B=0.32$ T and $T=1.25$ K\cite{si_p_t1}. For GaAs dots with
Fock-Darwin radius $l_{0}=30$ nm  and B=1 T one obtains
$T_{1}^{-1}\approx 200$ s$^{-1}$\cite{khaetskii}. These spin-flip
rates when compared to other decoherence mechanisms considered below
give negligible contributions,  with the possible exception of very
large dots ($l_{0}>100$ nm), due to the fact that $T_{1}^{-1}\propto
l_{0}^{8}$, a case we  do not consider here since reasonable
semiconductor QC architectures are limited by an inter-dot distance of
50 nm, necessary for exchange gate operations to work\cite{loss1}. 

We now describe the spin decoherence mechanisms of importance to QC
architectures.  Two electron spins may flip-flop  due to their dipolar
interaction. But this event is limited by inhomogeneous broadening,
because only spins with the same Zeeman splitting can satisfy energy
conservation  in this process\cite{note1}. The rate is given
by\cite{mims}
\begin{equation}
\frac{1}{T_{f}} \simeq \pi f(\omega_{0})\times \frac{\langle
\Delta\omega^{2}\rangle }{9}\simeq 0.33\frac{\langle \Delta\omega^{2}
\rangle}{(\Delta \omega)_{I}}.
\label{tf_rate}
\end{equation}
Here $f(\omega_{0})$ is the inhomogeneously broadened Gaussian
line-shape, which plays the role of the density of  states per unit
frequency, $(\Delta \omega)_{I}$ the inhomogeneous line-width,  while
$\langle \Delta \omega^{2} \rangle/9$ is the flip-flop contribution to
the second  moment\cite{mr_books}, which is of the order of the
transition matrix element squared.    It is important to mention that
the qubit-qubit dipolar interaction can in principle be included in
the Hamiltonian responsible for the quantum algorithm\cite{tahan02},
and also may be eliminated using magnetic resonance
techniques\cite{mr_books}. But here we have two reasons to include it
in our calculation: First we do not expect forthcoming $T_{M}$
measurements to be free from these effects. Second, since correcting
dipolar coupling implies additional overhead in QC design, it is
interesting to access the amount of error involved by ignoring its
presence.  Another spin-spin mechanism is spectral diffusion,  that
happens when the electron spins that generate the echo pulse  are
subject to fluctuating dipolar or hyperfine fields generated by
nuclear spins. A stochastic theory of this effect can be formulated by
treating the electron spin Larmor frequency $\omega$  as a random
variable, and calculating  the echo envelope amplitude $M(2\tau)$
using an ensemble average\cite{klauder}. The result is 
\begin{equation}
M(2\tau)\approx
M(0)\exp{\left[-(R\delta_{A}^{2}/12)(2\tau)^{3}\right]},
\label{decaygauss}
\end{equation}
where $\tau$ is the time interval between the $\pi/2$ and $\pi$
pulses, and $R$  the local field relaxation to a Gaussian probability
distribution with  width $\delta_{A}$. Hence the spectral diffusion
(SD) rate is given by  $T_{SD}^{-1}=(R\delta_{A}^{2}/12)^{1/3}$. The
decoherence rate follows from all the above contributions by solving
the cubic equation for $T_{M}^{-1}$:
$(T_{1}^{-1}+T_{f}^{-1})T_{M}+(T_{SD}^{-1}T_{M})^{3}=1$, and often one
finds it to be dominated by a single mechanism. (We give our
calculated values for $T_{M}$ in Si:P and GaAs quantum dot system in
Table 1, where the dominant decoherence mechanism in each case is also
listed.)

Now we estimate $T_{M}$ for Si:P.  For the dipolar mechanism, assuming
a qubit separation $d=10$ nm we get  $T_{f}^{-1}=1\times 10^{3}$
s$^{-1}$ and $2\times 10^{3}$ s$^{-1}$ when the qubits are arranged in
1D and 2D  lattices respectively. This is the case for $^{29}$Si
natural abundance ($f=4.67$\%) that leads to an experimental
line-width of 2.5 G\cite{feher1}.  However, for isotopically  pure
$^{28}$Si (which has zero spin), we expect no inhomogeneous broadening
[$(\Delta \omega)_{I}\sim \sqrt{\langle \Delta \omega^{2}\rangle}$ in
Eq.~(\ref{tf_rate})]. In that case, $T_{f}^{-1}$ will be much higher:
$T_{f}^{-1}=(2\langle \Delta \omega^{2}\rangle/\pi)^{1/2}=3\times
10^{5}$ s$^{-1}$, $4\times 10^{5}$ s$^{-1}$, dominated by qubit
dipolar coupling. The SD rate was measured for natural silicon by
Chiba and Hirai\cite{chiba}. By assuming that fluctuating dipolar
fields of $^{29}$Si nuclear spins caused Gaussian spectral diffusion
they formulated a theory to calculate the coefficient
$R\delta_{A}^{2}$ in (\ref{decaygauss}), which agreed within order of
magnitude with the measured $T_{M}$ (note that since their P
concentration was low $T_{f}^{-1}$ was negligible).  Therefore we
assume their experimental value of $T_{SD}^{-1}=3.3\times 10^{3}$
s$^{-1}$ as a reliable estimate for the SD rate for a single P spin in
a Si matrix, and by adding the dipolar flip-flop rate we are able to
get a phenomenological estimate for the decoherence rate when the P
spins are arranged in a quantum computer geometry, as opposed to a
dilute randomly doped sample. Hence we get $T_{M}^{-1}\approx 4\times
10^{3}$ s$^{-1}$ in both 1D and 2D geometries, noting that here both
spin-spin mechanisms contribute with the same order of magnitude. An
interesting consequence of the interplay of these two mechanisms is
that $T_{M}^{-1}$ displays a minimum as a function of the $^{29}$Si
fraction $f$, an effect not yet noted in the literature (Fig.~1). The
$T_{SD}^{-1}$ contribution is proportional to $f^{2/3}$, since the
probability of finding a pair of $^{29}$Si [analogous to $p_{ij}$ in
(\ref{spec_diff_rate}) below] is proportional to $f^{2}$, and then we
take a cubic root due to the non-exponential decay in
(\ref{decaygauss}). However, the dipolar flip-flop rate is decreased
when we increase $f$:  $T_{f}^{-1}\propto (\Delta
\omega)_{I}^{-1}\propto f^{-1/2}$. This happens because the $^{29}$Si
is the source of hyperfine broadening and the mean square deviation of
the hyperfine field is proportional to $f$. It would be interesting to
fit these results to measurements on isotopically purified samples; to
date there is only one measurement, with $f=0.12$\%, in which
$T_{M}^{-1}$ decreased by a factor of 2\cite{gordon} (the P
concentration was the same as in Fig.~1, $c=4\times 10^{16}$
cm$^{-3}$).  This is consistent with our result. For the case of a 2D
Si:P QC we find that $T_{M}^{-1}$ will attain a minimum when $f\approx
2$\%, suggesting that natural silicon is a good choice for QC
architectures. 

Turning to the GaAs-QD, we note that inhomogeneous broadening should
be much stronger: We estimate a line-width of at least 50 G due to
hyperfine interactions, since all nuclei have spin $3/2$ (S impurities
in GaAs have 500 G of broadening\cite{seck}). This together with the
fact that the qubits will be much further apart ($d=50$ nm) leads to
$T_{f}^{-1}<10^{-3}$ s$^{-1}$ ($T_{f}^{-1}\propto d^{-6}$).  We now
calculate the SD rate for these spins.  NMR in GaAs reveals a
composition of $^{75}$As (50\%), $^{69}$Ga (30.2\%), and $^{71}$Ga
(19.8\%)\cite{handbook}. The homogeneous line-width of the $^{75}$As
system (fcc lattice with $a_{0}=5.65$ \AA) is $\delta_{B}= 5.56\times
10^{3}$ s$^{-1}$\cite{shulman}, indicating that these nuclei are
flip-flopping every 300 $\mu$s. Since the hyperfine interaction
depends on the position of each nucleus, the electron feels a
different field if a pair of nuclei is up-down as opposed to down-up
(here we will neglect the contribution of nuclei outside the QD wave
function; they would produce a dipolar field which shifts the electron
frequency by at most $\delta \omega \sim
\gamma_{e}\gamma_{n}\hbar/z_{0}^{3}\sim 20$ s$^{-1}\ll
\delta_{r}=T_{1}^{-1}\sim 200$ s$^{-1}$, the intrinsic line-width due
to QD spin relaxation mentioned above; this is certainly not the case
in Si:P, where the nuclei outside the donor wave function give the
dominant contribution\cite{chiba}).  To calculate the spectral
diffusion rate we consider a three spin Hamiltonian $H=H_{0}+V$, where
\begin{eqnarray}
H_{0} &=& \hbar \omega_{i} I_{iz}S_{z}+\hbar \omega_{j}I_{jz}S_{z},
\label{h0}\\
V&=&\hbar b_{ij} (I_{i+}I_{j-}+I_{i-}I_{j+}) -4 \hbar b_{ij}
I_{iz}I_{jz},
\label{v}\\
b_{ij}&=&-\frac{1}{4}\gamma_{n}^{2}\hbar
\frac{1-3\cos^{2}{\theta_{ij}}}{R_{ij}^{3}},
\label{bij}\\
\omega_{i}&=& \omega_{h}
\exp{\left(-\frac{X_{i}^{2}+Y_{i}^{2}}{l_{0}^{2}}\right)}\cos^{2}{\left(
\pi \frac{Z_{i}}{z_{0}}\right)}. 
\label{omegai}
\end{eqnarray}
Here $H_{0}$ is the diagonal part of the hyperfine interaction (the
off-diagonal component can be neglected since electron-nucleus
flip-flop is forbidden by energy conservation), while $V$ is the
dipolar coupling between two nuclei. ${\mathbf I_{i}}$ is the spin
operator for the nucleus located at ${\mathbf
R_{i}}=(X_{i},Y_{i},Z_{i})$, $\gamma_{n} = 4.58 \times 10^{3}$ (s
G)$^{-1}$ the gyro-magnetic ratio for $^{75}$As, $R_{ij}$ the distance
between the two flip-flopping nuclei and $\theta_{ij}$ the angle
between this vector and the magnetic field. The hyperfine frequency is
given by 
\begin{equation}
\omega_{h} = \frac{8\pi}{3}\gamma_{e}\gamma_{n}\hbar \left| \Psi(0)
\right|^{2}= \frac{16}{3} \frac{\gamma_{e}\gamma_{n}\hbar
a_{0}^{3}}{z_{0}l_{0}^{2}}d(As), 
\end{equation}
where $d(As)=|u'(0)|^{2}$ is the electronic density on the nuclei and
$l_{0}$ the Fock-Darwin length.  The electron frequency change due to
a nuclear flip-flop is then $\Omega_{ij}=\omega_{i}-\omega_{j}$, while
the energy changes  by $\hbar\Omega_{ij}/2$ in this process. Since
$|b_{ij}|\lesssim |\Omega_{ij}|$ we can apply perturbation theory, and
the rate becomes\cite{chiba}
\begin{eqnarray}
\Gamma_{ij}&=&2\pi b_{ij}^{2}g_{B}(\Omega_{ij}/2),
\label{rate}\\
g_{B}(\omega)&=&\frac{1}{\sqrt{2\pi}\delta_{B}}\exp{\left(
-\frac{\omega^{2}}{2\delta_{B}^{2}}\right)}.
\label{bline}
\end{eqnarray}
The line-width $\delta_{B}$ of the nuclear system enters to guarantee
energy conservation: the change in the Zeeman energy of the electron
is compensated by the spin-spin interaction of the nuclear system. The
SD rate can now be estimated by summing over all pairs $ij$, as long
as they can flip-flop:
\begin{eqnarray}
\frac{1}{12}R\delta_{A}^{2}&=&
\frac{1}{12}\sum_{i<j}\Gamma_{ij}\Omega_{ij}^{2}
\left[1-\frac{f_{A}(\Omega_{ij})}{f_{A}(0)}\right]p_{ij}(T),
\label{spec_diff_rate}\\
f_{A}(\omega)&=&\frac{1}{\sqrt{2\pi}\delta_{r}}\exp{\left(
-\frac{\omega^{2}}{2\delta_{r}^{2}}\right)}.
\end{eqnarray}
Assuming the nuclei are in thermal equilibrium we get 
\begin{equation}
p_{ij}(T)= \left(
1+\frac{\cosh{3x}}{2\cosh{2x}+3\cosh{x}+2}\right)^{-1},
\label{pijT}
\end{equation}
with $x=\hbar \gamma_{n}B/k_{B}T$. For $B = 1$ T and  $T\gg 1$ mK we
assume the high $T$ limit $p_{ij}\approx 7/8$. By performing the sum
(\ref{spec_diff_rate}) numerically, we estimate $T_{SD}^{-1}\sim
10^{4}$ s$^{-1}$, that shows for QD's with $l_{0}\lesssim 100$ nm
spectral diffusion due to hyperfine field fluctuation dominates.  This
rate depends on B field intensity (Fig.~2), since $l_{0}$ decreases
when $B$ is increased, and also varies by about a factor of 2 when the
tilting angle with $z$ direction is changed [see Eq.~(\ref{bij})].

Finally, we wish to comment on the validity of the approximations
employed here.  The SD decay given by Eq. (\ref{decaygauss}) is only
valid for $R\tau \ll 1$\cite{klauder}, and also
Eq. (\ref{spec_diff_rate}) assumes a generalization of this two
parameter model to several parameters $\Gamma_{ij}$, $\Omega_{ij}$.
Certainly a more rigorous theory of SD due to nuclear spins needs to
be developed if one wants to go beyond the order of magnitude
estimates given here. In particular,  a rigorous theory for the
nuclear flip-flop rate [Eq. (\ref{rate})] is mandatory for a precise
description of this phenomena.  Our estimated coherence times,
summarized in Table 1, should be compared with the longest gating time
in the corresponding QC architecture. For the parameters chosen here,
the exchange time will be $\tau_{J}\sim \hbar/0.1$ meV $\sim 1$ ps,
with a typical exchange coupling\cite{loss1} of 0.1 meV. A single
qubit rotation (Rabi flop) can be done with an ESR field 20 times
smaller than the applied field, that leads to a $\pi/2$ rotation time
$\tau_{R}\sim 20/\gamma B \sim 0.1$ ns, $0.5$ ns at $B=1$ T for Si:P,
GaAs-QD respectively.  These time scales lead to a quality factor
$T_{M}/\tau_{R}>10^{4}$ for both architectures.  Also,
$T_{M}/\tau_{J}\sim 10^{6}-10^{8}$ (since $\tau_{J}\sim 1$ ps and
$T_{M}\sim 1-100$ $\mu$s), which implies a large number ($\gg 10^{4}$)
of coherent  gate operations allowing convenient fault tolerant
computation well within the currently estimated ($10^{-4}$) error
correction scheme.  Hence the electron spin in semiconductors is
confirmed as a competitive qubit candidate, with the effective low
temperature upper bound on the coherence time given by $1-100$ $\mu$s
under quite general conditions. The authors acknowledge discussions
with J. Fabian, X. Hu, A. Kaminski, B. Koiller, and I. \v{Z}uti\'{c}.
This work is supported by ARDA, LPS, US-ONR, and NSF.

\begin{table}
\begin{tabular}{|c|cc|cc|cc|}
Architecture & $T_{M}$ [$\mu$s] && $Q=T_{M}/\tau_{R}$ && Dominant
Mech. &  \\ \hline \hline Si:P/natural Si & 200 && $10^{6}$ &&
Dip./Spec. Diff. &\\ \hline Si:P/pure $^{28}$Si & 2 && $10^{4}$ &&
Dipolar &\\ \hline GaAs-QD & 50 && $10^{5}$ && Spec. Diff.&\\ 
\end{tabular}
\caption{ Coherence times $T_{M}$, quality factors, and dominant
decoherence mechanisms  for three quantum computer architectures.  We
assume the qubits are disposed in a 2D square lattice of side 10 nm and
50 nm for the  case of Si:P and GaAs-QD  respectively. For the GaAs-QD
architecture, Fock-Darwin radius is assumed 30 nm.}   
\end{table}

\begin{figure}

\epsfysize=6.0cm \centerline{\epsffile{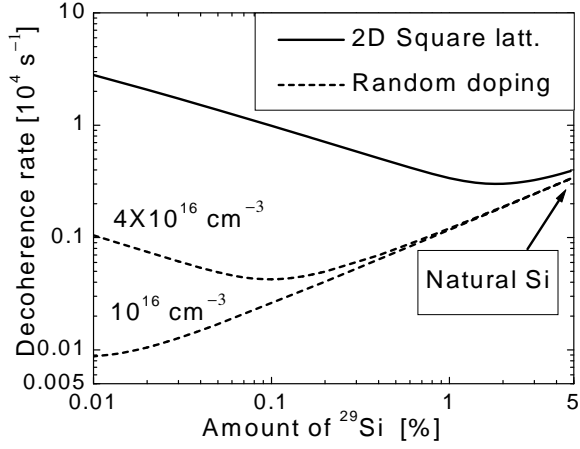}}
\vspace{0.5cm}
\caption{   Depicts the competition between the dipolar flip-flop rate
and spectral diffusion for a bound-electron spin in Si:P, leading to a
minimum in the decoherence rate as a function of the $^{29}$Si
fraction. We show calculations for a 2D quantum computer architecture
with qubit separation of 10nm, and for the most common experimental
situation of random phosphorus doping, with concentrations $1-4\times
10^{16}$ cm$^{-3}$.
\label{figone}}
\end{figure}

\begin{figure}

\epsfysize=6.0cm \centerline{\epsffile{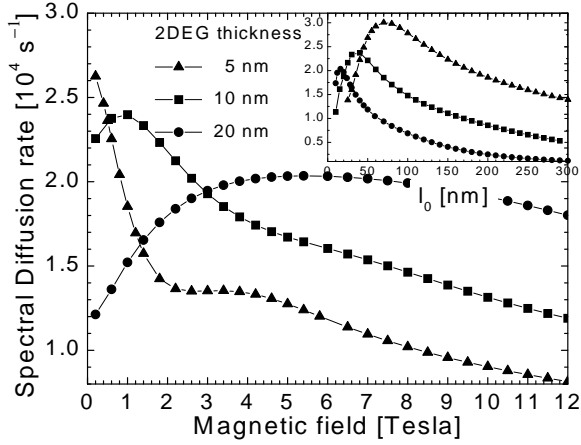}}
\vspace{0.5cm}
\caption{ Calculated spectral diffusion rate as a function of magnetic field and
quantum dot Fock-Darwin radius $l_{0}$ (inset).  For the B field plot
the dot transverse confinement length is set to 50 nm. The spectral
diffusion mechanism completely dominates the decoherence rate for
small dots ($l_{0}<$100 nm).
\label{figtwo}}
\end{figure}


\begin{thebibliography}{99}
\bibitem{note1} Flip-flop followed by phonon emission in an
inhomogeneously broadened line is analogous to phonon emission
mediated by hyperfine interaction. This rate is
proportional to $(\Delta H)_{I}^{3}$, and is negligible in our cases
($\Gamma < 10^{-15}$ s$^{-1}$).
\end{thebibliography}
\end{document}